\documentclass{article}

\usepackage{arxiv}

\usepackage[utf8]{inputenc} % allow utf-8 input
\usepackage[T1]{fontenc}    % use 8-bit T1 fonts
\usepackage{hyperref}       % hyperlinks
\usepackage{url}            % simple URL typesetting
\usepackage{booktabs}       % professional-quality tables
\usepackage{amsfonts}       % blackboard math symbols
\usepackage{nicefrac}       % compact symbols for 1/2, etc.
\usepackage{microtype}  % microtypography
\usepackage{lipsum}		% Can be removed after putting your text content
\usepackage{graphicx}
\usepackage[numbers]{natbib}
\usepackage{doi}

\title{Never a Dull Moment: Distributional Properties as a Baseline for Time-Series Classification}
\author{Trent Henderson\\ The University of Sydney
\And Annie G. Bryant\\ The University of Sydney
\And Ben D. Fulcher\\ The University of Sydney}

\renewcommand{\headeright}{}
\renewcommand{\undertitle}{}
\renewcommand{\shorttitle}{Never a Dull Moment}

\begin{document}
\maketitle

\begin{abstract}
The variety of complex algorithmic approaches for tackling time-series classification problems has grown considerably over the past decades, including the development of sophisticated but challenging-to-interpret deep-learning-based methods.
But without comparison to simpler methods it can be difficult to determine when such complexity is required to obtain strong performance on a given problem.
Here we evaluate the performance of an extremely simple classification approach---a linear classifier in the space of two simple features that ignore the sequential ordering of the data: the mean and standard deviation of time-series values.
Across a large repository of 128 univariate time-series classification problems, this simple distributional moment-based approach outperformed chance on 69 problems, and reached 100\% accuracy on two problems.
With a neuroimaging time-series case study, we find that a simple linear model based on the mean and standard deviation performs better at classifying individuals with schizophrenia than a model that additionally includes features of the time-series dynamics.
Comparing the performance of simple distributional features of a time series provides important context for interpreting the performance of complex time-series classification models, which may not always be required to obtain high accuracy.

\end{abstract}

\section{Introduction}

Time-series classification is a key problem in the sciences and industry wherein time-varying data is used to distinguish labeled classes.
The quantity and diversity of time-series classification algorithms is large and increasing, from simple linear decision boundaries in interpretable feature spaces \cite{fulcherHighlyComparativeFeaturebased2014} to complex methods based on deep neural networks, such as long short-term memory networks \cite{ismailfawazDeepLearningTime2019}.
Complex new algorithms can yield impressive classification accuracy on challenging problems, but often at the expense of clear human interpretability \cite{rudin2019stop}.

The UEA/UCR univariate time-series classification repository, which currently contains 128 problems spanning a variety of domains \cite{UEAUCRRepository}, has been crucial for encouraging transparent reporting of the relative strengths and weaknesses of classification algorithms.
It has also enabled the field to overcome key limitations in prior standard practice of reporting and comparison, including avoiding cherry-picking of optimistic datasets when reporting algorithm performance \cite{Keogh2003:NeedTimeSeries}.
Systematic comparisons of time-series classification algorithms across this database have been essential for benchmarking the accuracy of state-of-the-art algorithms \cite{bagnallGreatTimeSeries2017}.
However, in settings ranging from policy-making \cite{goodman2017european} to healthcare \cite{caruana2015intelligible}, deriving interpretable understanding that can guide subsequent decision-making can be more important than raw classification accuracy.
In such settings, simpler methods that are faster to train and clearer to interpret, are often preferred.

Recent work has shown that parsimonious and interpretable methods can match (or even outperform) more sophisticated ones, in settings ranging from sleep-stage classification \cite{van2023not} to earthquake detection \cite{eage:/content/papers/10.3997/2214-4609.202011128}.
A particularly striking example is the recent demonstration of a two-parameter logistic regression model with equivalent performance in earthquake aftershock forecasting to a deep neural network with thousands of parameters \cite{mignan_one_2019}.
For such settings, in which simple methods can perform well, the choice to instead use overly complex models risks over-fitting, leading to poor generalizability on out-of-sample data.
This effect has been demonstrated in a recent meta-analysis of deep-learning models for classifying autism spectrum disorder from resting-state neuroimaging data, which exhibited inferior performance to linear support vector machine (SVM) models on unseen data \cite{traut2022insights}.

Defaulting to complex classification models and comparing their performance to chance-level accuracy---or even other complex methods---can thus result in classifiers that are over-complicated and difficult to interpret.
This issue is well-illustrated by the time-series classification task of distinguishing `epilepsy' from `eyes-open' states from electroencelphalogram (EEG) data \cite{Andrzejak2001:IndicationsNonlinearDeterministic}.
While state-of-the-art approaches had applied complex algorithmic approaches, ranging from independent components analysis and discrete wavelet transforms to multi-layer neural networks \cite{Subasi2010:EEGSignalClassification}, it was found that the time-series standard deviation alone could completely separate the two classes \cite{fulcherHighlyComparativeTimeseries2013}.
The strong performance of a simple threshold classifier on standard deviation thus suggests it as an interpretable and parsimonious alternative to overly complicated algorithmic approaches to this problem, highlighting the utility of starting with a simple and parsimonious model and building in complexity only when it yields clearly demonstrable benefits.

Given that the characteristic complexity of time-series classification tasks (compared to classification using non-sequential data) relates to the challenge of quantifying class-informative dynamical patterns, here we aimed to investigate the performance of an extremely simple benchmark: a linear classifier in the two-dimensional space of two extracted features that ignore the sequential ordering entirely: the mean and standard deviation.
As these features characterize the distribution of time-series values (and are thus unrelated to the sequential ordering), high performance of this simple benchmark indicates problems for which trivial properties of the distribution are already sufficient to perform well, undermining the need for more complex approaches that aim to quantify informative temporal patterns \cite{fulcherFeatureBasedTimeSeriesAnalysis2018}.
While some previous work \cite{fulcherHighlyComparativeFeaturebased2014,lubbaCatch22CAnonicalTimeseries2019} has $z$-scored all time series prior to analysis---thereby insuring mean and variance are uninformative, to focus on dynamical patterns---such normalization has not been applied consistently across all problems in the UEA/UCR Repository \cite{bagnallGreatTimeSeries2017}.
We also investigate the performance gains of additionally incorporating a simple set of time-series features that capture different types of dynamical patterns in the data, using the \textit{catch22} feature set \cite{lubbaCatch22CAnonicalTimeseries2019}.
None of these features are sensitive to the mean or standard deviation of the data; they instead capture more subtle properties of the time series, including properties of distributional shape; periodic patterns; temporal spacing of outliers; and linear and nonlinear autocorrelation \cite{lubbaCatch22CAnonicalTimeseries2019}.
Finally, to demonstrate the practical implications of our findings, we present a case study of classifying individuals with schizophrenia using functional neuroimaging time series.

\section{Methods}

This study aims to evaluate the performance of a simple baseline linear classifier for time series in the two-dimensional space of their mean ($\mu$) and standard deviation ($\sigma$)---features related to the First Two Moments (FTM) of the distribution.
We first discuss methods related to fitting the FTM and \textit{catch22} feature-based time-series classifiers across the UEA/UCR Repository \cite{UEAUCRRepository} (in Sec.~\ref{sec:methods_benchmark}) and then describe specific methods related to our neuroimaging case study (in Sec.~\ref{sec:methods_scz}).
Code for reproducing our results is available on GitHub\footnote{\url{https://github.com/hendersontrent/mean-var-ts-classify}}.

\subsection{Feature-based classifier performance on the UEA/UCR Repository}
\label{sec:methods_benchmark}

In the UEA/UCR Time-Series Classification Repository \cite{UEAUCRRepository}, each of the 128 problems (as downloaded on 9 Feb 2023) are partitioned into designated train and test sets.
These problems vary widely in the number of time series, the relative size of train and test set sizes, time-series lengths, and number of classes (see \cite{UEAUCRRepository,bagnallGreatTimeSeries2017} for details).
For all time series, we computed the FTM feature set containing two features: the mean ($\mu)$ and standard deviation ($\sigma$).
For comparison to more sophisticated features of the time-series dynamics, we used the \textit{catch22} set of 22 time-series features \cite{lubbaCatch22CAnonicalTimeseries2019}.
Given a feature space, we fit and evaluated classification models following the resample-based procedure outlined in \cite{bagnallGreatTimeSeries2017}, using 30 resamples of train--test splits.
In addition to the designated split, we generated 29 additional (seeded) resamples of the data that preserve the class proportions of the designated split, for a total of 30 train--test splits.
Prior to fitting a linear SVM, features were normalized as a $z$-score by computing the mean and standard deviation of each feature in the train set, and using these values to rescale both the train and test sets (ensuring that test data was completely unseen).
A linear SVM was fit on the train set and used to generate predictions for the test set.
Classification accuracy was used as the performance metric, following prior work.
To statistically compare the performance of FTM, and the union set of 24 features from both \textit{catch22} and FTM (\textit{catch22} + FTM), we implemented a correlated test statistic \cite{nadeauInferenceGeneralizationError2003} that corrects the traditional $T$-statistic to account for the violation of the assumption of independence incurred in the usage of resampling using the \textit{correctR} package \cite{correctR}.
To compare the FTM and \textit{catch22} + FTM feature sets, we removed four problems (\texttt{AllGestureWiimoteX}, \texttt{AllGestureWiimoteY}, \texttt{AllGestureWiimoteX}, and \texttt{PLAID}) due to presence of time series that had $< 10$ real values before containing all or mostly missing values---the minimum threshold for calculations in \textit{catch22}.

\subsection{Case study: Schizophrenia classification}
\label{sec:methods_scz}

As a case study, we investigated the performance of simple feature-based classifiers to distinguish adults with schizophrenia versus cognitively healthy controls based on their whole-brain activity dynamics.
We obtained resting-state functional magnetic resonance imaging (rs-fMRI) data from the University of California at Los Angeles Consortium for Neuropsychiatric Phenomics LA5c Study \cite{poldrack2016phenome}, which was pre-processed using the ICA-AROMA + 2P + GMR method, as described previously \cite{aquino2020identifying}.
Blood oxygen level-dependent (BOLD) signals consisting of 152 time samples were extracted for each of 68 cortical \cite{desikan2006automated} and 14 subcortical \cite{fischl2002whole} regions per participant.
Any participants with flatline time series across all brain regions after preprocessing were excluded, yielding a final sample of $N = 166$ participants ($N = 118$ control and $N = 48$ schizophrenia). 
The two groups differ in terms of age (Control $= 31.5 \pm 8.8$ years; Schizophrenia $= 36.6 \pm 9.0$ years; Wilcoxon rank sum test $p < 0.01$) and sex (Control $= 46.6\%$ Female; Schizophrenia = $25\%$ Female; $X^2 (1, N = 166) = 5.75$, $p < 0.05$).

After feature extraction (using \textit{theft} \cite{hendersonFeatureBasedTimeSeriesAnalysis2022}), the dataset was in the form of an $N \times R \times F$ matrix, for $N$ subjects, $R$ brain regions, and $F$ features.
We incorporated all combinations of $82$ brain regions and features (either $2$ FTM features or $24$ FTM + \textit{catch22} features) as inputs to a regularized linear SVM classifier (from \textit{scikit-learn} \cite{scikit-learn}), setting the regularization parameter \texttt{C = 1}, disabling the shrinking heuristic, and using balanced class weights.
In other words, the FTM-only model had a total of $2 \times 82 = 164$ SVM input features and the FTM + \textit{catch22} model had a total of $24 \times 82 = 1968$ features.

Model performance was evaluated as balanced accuracy using $10$-fold cross-validation (CV) with 10 repeats (setting the random state such that all compared models received the same resampled sets of test folds).
For each $k$-fold, the same $z$-score feature-normalization was applied as described in Sec.~\ref{sec:methods_benchmark}.
Balanced accuracy is reported as the mean $\pm$ SD across the 10 repeats, in which each repeat contains the mean balanced accuracy across the 10 CV folds.
Statistical significance of a given balanced accuracy value was assessed using permutation testing using 1000 null samples obtained from shuffling class labels, pooling null samples across all combinations of regions and features for each feature set \cite{leek2011joint}.
False positives were controlled at the $\alpha = 0.05$ level using Bonferroni correction.

\section{Results}

Results are structured as follows.
First, in Sec.~\ref{sec:UCR_results}, we describe findings across the UEA/UCR Repository, where we demonstrate that the FTM classifier outperforms chance on the majority of problems.
We then highlight the practical ramifications of these results using the example of a neuroimaging biomarker classification task in Sec.~\ref{sec:results_scz}, in which the mean and standard deviation exhibit strong performance that is weakened by adding dynamical properties of the functional neuroimaging time series.

\subsection{Linear FTM classifier performance across the UEA/UCR Database}
\label{sec:UCR_results}

\begin{figure}[tbph!]
  \centering
  \includegraphics[width = 0.86\textwidth]{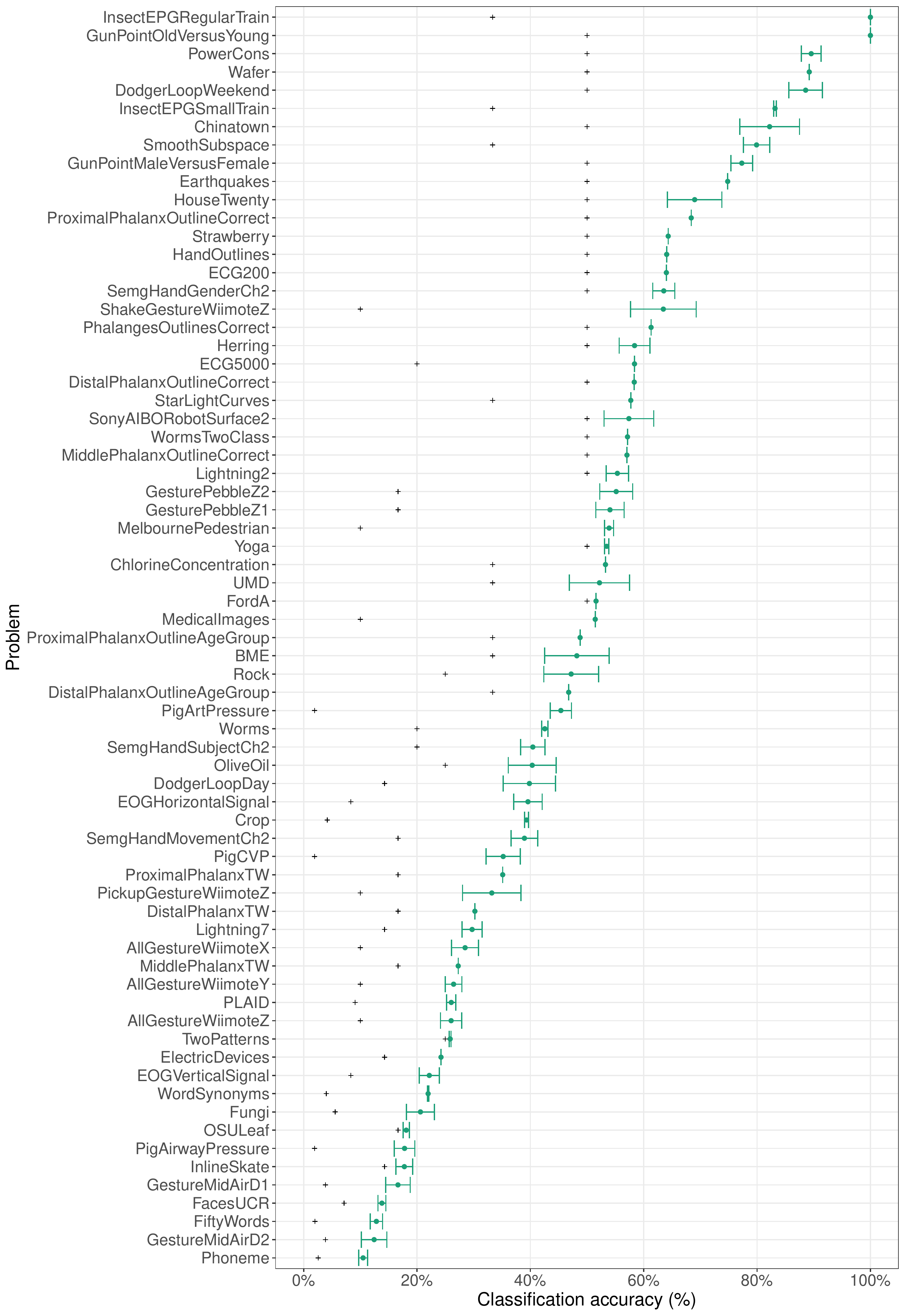}
  \caption{\label{fig:meanvarplot}
  \textbf{Mean and standard deviation as features in a linear SVM classifier statistically outperforms chance-level accuracy on 69 of 128 problems in the UEA/UCR time-series repository.}
Classification accuracy ($\%$) is displayed along the horizontal axis and the name of each problem is shown on the vertical axis.
Chance accuracies are displayed as black crosses.
Points indicate mean classification accuracy and error bars show $\pm 1$ standard deviation (across train--test resamples).
  }
\end{figure}

We first investigated how a linear classifier based on the two FTM features, $(\mu,\sigma)$, performs across the 128 problems from the UEA/UCR Repository.
To determine the problems on which FTM beats chance, we calculated a $p$-value of the chance probability against the distribution of 30 resampled FTM accuracy values for each problem.
We found that the FTM-based classifier statistically outperformed chance ($p < 0.05$) on 69 of the 128 problems (i.e., on $53.9\%$ of problems).
FTM-based classification accuracies relative to chance level are plotted for these 69 problems in Fig.~\ref{fig:meanvarplot}.
We note that chance is often beaten by a substantial margin on these problems.
Indeed, this simple classifier achieves 100\% accuracy on two problems: \texttt{InsectEPGRegularTrain} and \texttt{GunPointOldVersusYoung}.
These results demonstrate that, even on a repository devoted to time-series classification tasks, strong performance can be obtained using the two simplest distributional statistics that are unrelated to the sequential ordering of the data, due to the database containing problems in which labeled classes have distinctive levels (means) or scales (variances).

\begin{figure}[h]
  \centering
  \includegraphics[width = 0.95\textwidth]{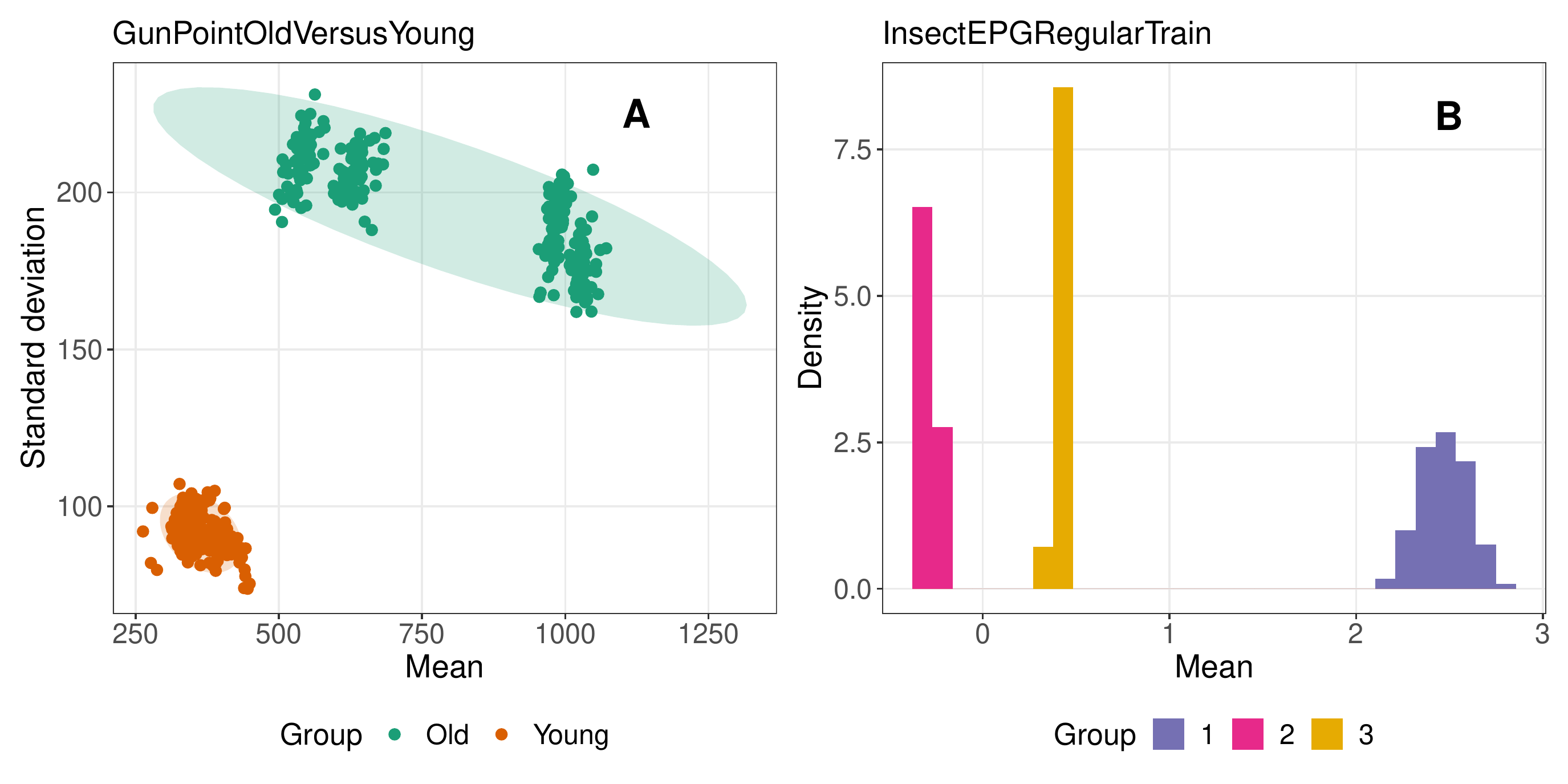}
  \caption{\label{fig:meanvarcaseplot}
  \textbf{Simple distributional moments can perfectly separate classes in the \texttt{GunPointOldVersusYoung} and  \texttt{InsectEPGRegularTrain} datasets.}
\textbf{A} Individual time series in \texttt{GunPointOldVersusYoung} are represented in the two-dimensional feature space of mean and standard deviation.
Points are colored by class (as labeled) and class-level covariances are shown as shaded ellipses to guide the eye.
\textbf{B} Histogram of time-series means by class in \texttt{InsectEPGRegularTrain}.
  }
\end{figure}

To better understand how the FTM classifier is behaving, we examined the two problems for which it achieved 100\% accuracy.
For \texttt{GunPointOldVersusYoung}, the classes are clearly distinguished in the $(\mu,\sigma)$ feature space (Fig.~\ref{fig:meanvarcaseplot}A), while for \texttt{InsectEPGRegularTrain} each class has a characteristic mean level (Fig.~\ref{fig:meanvarcaseplot}B).
For the binary (`Young' vs `Old') \texttt{GunPointOldVersusYoung} task, the time series is the $x$-axis coordinate of their centre of the hand at each frame while moving it from rest position to a gun pose and back again.
Figure~\ref{fig:meanvarcaseplot}A shows that the `Young' actor exhibits characteristically lower mean and variation of their hand coordinate, consistent with them being shorter and having a shorter arm than the `Old' actor.
It is worth noting that, in contrast to features of the time-series dynamics (many of which are invariant to linear rescalings of the time-series values), $\mu$ and $\sigma$ are highly sensitive to the calibration of experimental measurement.
For the \texttt{GunPoint} collection of problems, this makes these features less likely to generalize well to new examples, with small differences in the angle of the camera used to measure hand coordinates, or the distance from camera to actor, strongly affecting differences in $\mu$ and $\sigma$ and hence accurate classification \cite{Wu2023:WhenEarlyClassification}.
In \texttt{InsectEPGRegularTrain}, each class has a characteristic mean voltage level of the electrical circuit that connects insects with their food source, allowing the classes to be accurately distinguished by ignoring sequential patterns, and simply focusing on this mean voltage.

Finally, we aimed to investigate the extent to which measuring the \textit{catch22} features in addition to the FTM features would improve performance.
We compared the performance of the FTM feature set (2 features) to that of the \textit{catch22} + FTM feature set (24 features) across the 124 problems where \textit{catch22} features could be successfully calculated (four problems were excluded due to a high number of missing values, cf. Methods).
Adding \textit{catch22} features resulted in an average absolute improvement in classification accuracy of $33.7\%$ across the 124 problems, confirming the general importance of capturing dynamical properties for time-series classification problems.
However, pairwise comparisons using the corrected test statistic revealed that there was no statistical difference between FTM and \textit{catch22} + FTM on $45$ of the $124$ (or $36.4\%$ of) problems.
This demonstrates that for many problems, simple distributional properties yield a surprisingly strong baseline against which to assess the benefits gained by using more complex approaches.

\subsection{Neuroimaging Biomarker Case Study}
\label{sec:results_scz}

\begin{figure}[h]
  \centering
  \includegraphics[width = 0.4\textwidth]{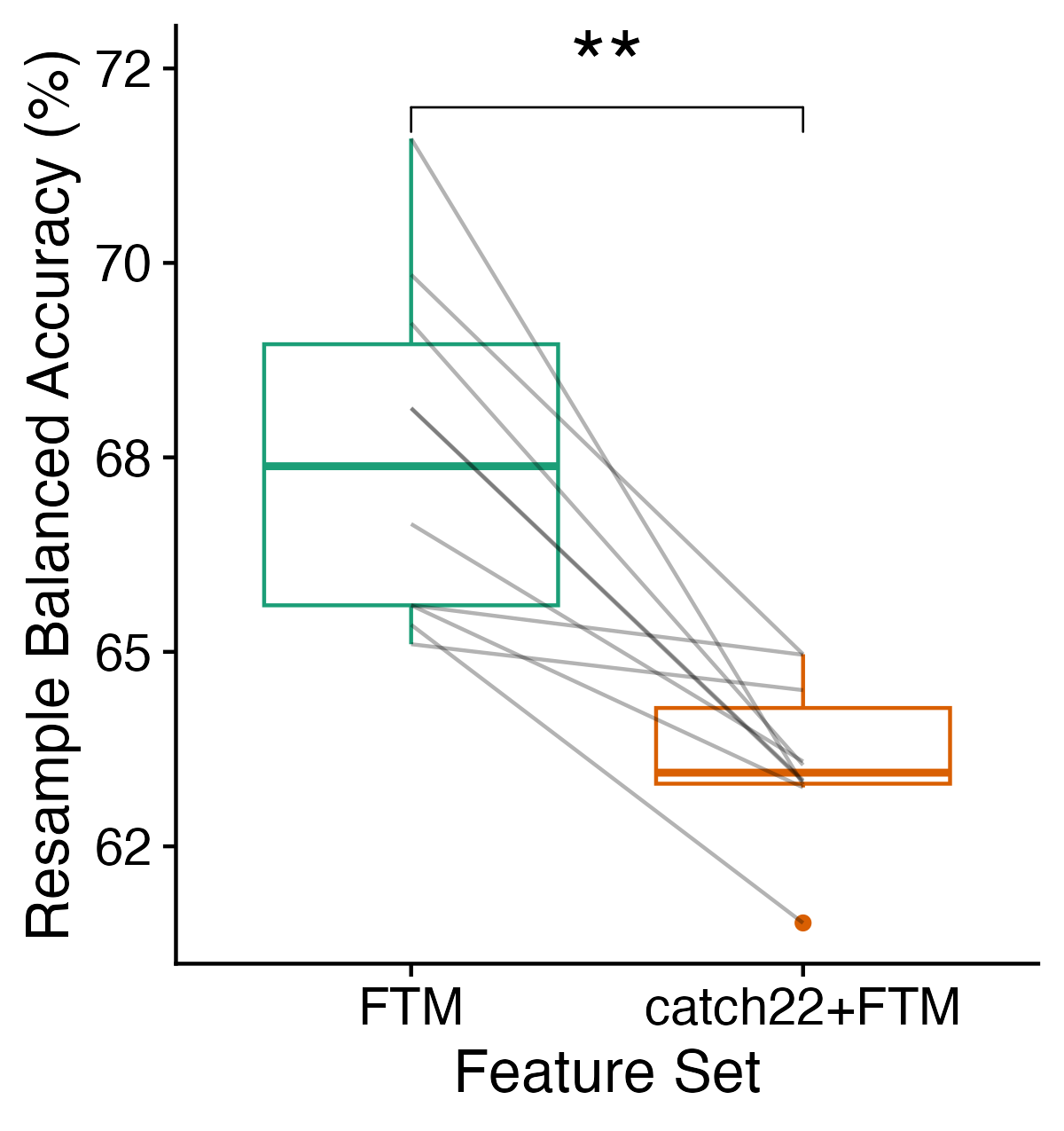}
  \caption{
  \label{fig:schizophrenia_FTM_catch24}
  \textbf{The FTM feature set shows strong performance at classifying individuals with schizophrenia from resting-state fMRI time series.}
The distribution of mean balanced accuracy (across 10 repeats), using the combination of all brain regions and either the two \textit{FTM} features (green), or the 24 FTM + \textit{catch22} features (orange), is shown as a combined spaghetti plot and boxplot.
Each gray line indicates the same resampled set of $k = 10$ folds evaluated for each feature set along the $x$-axis.
$^{\ast\ast}$: $p = 0.003$, corrected resampled $T$ $= 3.63$.
}
\end{figure}

We next extended our investigation of simple baseline time-series classifiers to a schizophrenia classification problem using rs-fMRI time series, a setting in which complex, opaque classifiers are common \cite{quaak2021deep,zeng2018multi}.
Given the strong performance of simple distributional statistics for some time-series classification problems above, we evaluated the performance of FTM alone (and with the addition of the \textit{catch22} feature set) in classifying individuals with schizophrenia from healthy controls.

As shown in Fig.~\ref{fig:schizophrenia_FTM_catch24}, the model based on FTM features displayed a high balanced accuracy ($67.5 \pm 2.2\%$), which sits well within the range of a recent meta-analysis of schizophrenia classification studies using rs-fMRI \cite{de2019machine}.
Adding \textit{catch22} features to the model decreased its performance to $63.6 \pm 1.0\%$ (corrected resampled $T$-statistic = $3.63$, $p = 0.003$).
Consistent with the findings on some problems in the UAE/UCR Repository above, these results demonstrate the surprisingly strong performance of basic properties of the distribution of time-series values for this fMRI classification task.
Surprisingly, for this rich and complex whole-brain time-series dataset, incorporating information about temporal patterns in the data (using \textit{catch22}) yields models with inferior performance than the simple two-dimensional model based on mean and standard deviation.

\section{Discussion}

With the growing sophistication of machine-learning algorithms, modern data analysts have increasingly adopted a methodological approach that defaults to more complex statistical methods, which can obscure their clear interpretation.
Presenting such complex approaches without direct comparison to simpler alternatives makes it difficult to discern whether the methodological complexity is beneficial relative to simpler approaches, an assumption that has been challenged in some recent studies \cite{van2023not,mignan_one_2019,eage:/content/papers/10.3997/2214-4609.202011128}.
In this work, focusing on time-series classification problems, we demonstrate that a method at the extreme end of simplicity (using just two distributional-moment-based features, $\mu$ and $\sigma$, and a linear classifier) performs surprisingly well on many such problems.
Despite being insensitive to the unique property of sequential data relative to an unordered vector---its ordering (in time)---this simple benchmark classifier statistically outperformed chance on approximately half of the problems in a prominent time-series classification archive, the UEA/UCR Repository \cite{UEAUCRRepository}, and even achieved $100\%$ accuracy on two problems.
Our results emphasize the importance of carefully considering the factors that contribute to model performance, assessing model gains relative to simpler models, and favoring parsimony by carefully building up model complexity incrementally from simple baselines.

We demonstrated the applicability of our findings to a time-series classification problem in neuroimaging, a setting in which many prior models have been highly complex models \cite{quaak2021deep} but can fail to generalize \cite{traut2022insights}.
Importantly, classification performance was stronger with FTM than with a model that also included \textit{catch22} features of the dynamics using an SVM classifier, which is generally the top-performing classifier type in rs-fMRI analysis for schizophrenia \cite{de2019machine}.
The combination of $\sigma$ and $\mu$ yielded a mean balanced accuracy of $67.5\%$ while affording clear statistical and biological interpretations.

The strong performance of mean and standard deviation across the UEA/UCR Repository demonstrates that time series in these problems have not been consistently normalized, as prior work has highlighted \cite{bagnallGreatTimeSeries2017}.
If all time series were individually $z$-score transformed, there would be no class differences in mean or standard deviation for any problem (and the FTM-based classifier would exhibit null performance).
Our results have implications for comparing time-series classification algorithms on problems for which labeled classes can be distinguished based on distributional properties alone.
For example, consider the \textit{catch22} features, which are all insensitive to the mean and variance of the input time series \cite{lubbaCatch22CAnonicalTimeseries2019}.
Relative to \textit{catch22}, the superior performance of an alternative time-series classifier (which \textit{is} sensitive to $\mu$ or $\sigma$ of the input series) could be driven entirely by a class-relevant difference in $\mu$ or $\sigma$: properties that are unrelated to dynamical patterns (and that \textit{catch22} cannot access).
One approach to testing the ability of different classification algorithms to capture properties related to the dynamics of a uniformly sampled univariate time series would be to normalize time series such that there are no class differences in basic distributional properties.
A second approach would be to compare model performance to a benchmark in which only simple distributional properties are included, with gains relative to this benchmark then being attributable to class-relevant differences in more complex properties.
In this latter approach, there is scope for extending the two simple (FTM) features used here by adding higher-order moments, or other types of distributional features (such as those included in \textit{hctsa} \cite{fulcherHctsaComputationalFramework2017}).

This work also highlights the need for careful consideration of the generalizability of models that use features sensitive to the calibration of time-series measurements (like $\mu$ and $\sigma$), relative to features that are invariant to linear rescalings of the input time series (like the \textit{catch22} feature set).
This is because such measurement-scale-dependent features are highly sensitive to changes in the calibration of experimental measurements that may not be precisely maintained in new data.
An illustrative example is shown here in the \texttt{GunPointOldVersusYoung} dataset, for which our FTM-based method achieved 100\% accuracy.
However, this may be a deceptively high value, as it is reliant on the precise calibration of new data.
A prior analysis of this dataset, focusing on early classification, has shown that classification accuracy can plummet with only slight variability in experimental calibration (to changes as small as a $\approx 1.9^\circ$ tilt in the angle of the camera used to measure hand coordinates) \cite{Wu2023:WhenEarlyClassification}.
In general, the decision of whether or not to include features that are sensitive to measurement scale (or focus on scale-invariant features of the dynamics or distribution shape, as in \textit{catch22}) should be motivated by domain expertise to avoid overly optimistic classification results.

In summary, by highlighting many time-series classification problems for which simple distributional properties of a time series can achieve surprisingly high classification accuracy, our results raise important issues for the development and interpretation of time-series classification models.
Future work on evaluating time-series classification algorithms may consider using simple benchmarks for comparison to aid interpretation and provide evidence for the contribution of model complexity to any performance advantage, particularly for problems highlighted here for which simple distributional features are highly informative of class differences.

\subsection*{Acknowledgements}
The authors would like to thank Kevin Aquino for sharing preprocessed fMRI data \cite{Aquino2022:IntersectionDataQuality} used in the case study, and Eamonn Keogh for providing useful feedback on a manuscript draft.

\bibliographystyle{benbibstyle}
\bibliography{references}  %%% Uncomment this line and comment out the ``thebibliography'' section below to use the external .bib file (using bibtex) .

%%% Uncomment this section and comment out the \bibliography{references} line above to use inline references.
% \begin{thebibliography}{1}

% 	\bibitem{kour2014real}
% 	George Kour and Raid Saabne.
% 	\newblock Real-time segmentation of on-line handwritten arabic script.
% 	\newblock In {\em Frontiers in Handwriting Recognition (ICFHR), 2014 14th
% 			International Conference on}, pages 417--422. IEEE, 2014.

% 	\bibitem{kour2014fast}
% 	George Kour and Raid Saabne.
% 	\newblock Fast classification of handwritten on-line arabic characters.
% 	\newblock In {\em Soft Computing and Pattern Recognition (SoCPaR), 2014 6th
% 			International Conference of}, pages 312--318. IEEE, 2014.

% 	\bibitem{hadash2018estimate}
% 	Guy Hadash, Einat Kermany, Boaz Carmeli, Ofer Lavi, George Kour, and Alon
% 	Jacovi.
% 	\newblock Estimate and replace: A novel approach to integrating deep neural
% 	networks with existing applications.
% 	\newblock {\em arXiv preprint arXiv:1804.09028}, 2018.

% \end{thebibliography}

\end{document}